\documentstyle[preprint,aps,eqsecnum]{revtex}

\def\beq#1{\begin{equation} \label{#1}}
\def\eeq{\end{equation}}
\def\bra#1{\left\langle #1\right\vert}
\def\ket#1{\left\vert #1\right\rangle}

\input prepictex
\input pictex
\input postpictex
\newdimen\tdim
\tdim=\unitlength
\def\stpltsmbl{\setplotsymbol ({\small .})}

\newbox\phru
\setbox\phru=\hbox{\beginpicture
\setcoordinatesystem units <\tdim,\tdim>
\stpltsmbl
\setquadratic
\plot
0 0
2.5 3
5 0
7.5 -3
10 0
/
\endpicture}
\def\photonru #1 #2 *#3 /{\multiput {\copy\phru}  at
#1 #2 *#3 10 0 /}

\newbox\sru
\setbox\sru=\hbox{\beginpicture
\setcoordinatesystem units <\tdim,\tdim>
\stpltsmbl
\setquadratic
\plot
  0.0   0.0
  4.8   1.5
  7.5   5.0
  7.3   8.5
  5.0  10.0
  2.7   8.5
  2.5   5.0
  5.2   1.5
 10.0   0.0
/
\endpicture}
\def\springru #1 #2 *#3 /{\multiput {\copy\sru}  at
#1 #2 *#3 10 0 /}

\def\beq#1{\begin{equation} \label{#1}}
\def\eeq{\end{equation}}
\def\bra#1{\left\langle #1\right\vert}
\def\ket#1{\left\vert #1\right\rangle}

\begin{document}
{
\tighten
%\preprint {\vbox{
% \hbox{WIS/14/06-SEPT-DPP
% \hbox{TAUP 2834/06}
% \hbox{ANL-HEP-PR-06-63}
% \hbox{hep-ph/0608284}
% \hbox{}
%}}
\begin{center}
{\Large\bf Pauli blocking and entanglement solve $K\pi$ puzzle\\
CP violation in
 $B^o \rightarrow K\pi$; not in $B^{\pm} \rightarrow K\pi$ decays}
\vrule height 2.5ex depth 0ex width 0pt
%\mydraft
\vskip0.8cm
Harry J. Lipkin\,$^{b,c}$\footnote{e-mail: \tt ftlipkin@weizmann.ac.il} \\
\vskip0.8cm
{\it
$^b\;$School of Physics and Astronomy \\
Raymond and Beverly Sackler Faculty of Exact Sciences \\
Tel Aviv University, Tel Aviv, Israel\\
\vbox{\vskip 0.0truecm}
$^c\;$Department of Particle Physics \\
Weizmann Institute of Science, Rehovot 76100, Israel \\
and\\
High Energy Physics Division, Argonne National Laboratory \\
Argonne, IL 60439-4815, USA
}
% end of \it
\end{center}

\vspace*{0.8cm}

\centerline{\bf Abstract}
New data analysis with Pauli blocking and entanglement explains CP violation in $B^o\rightarrow K\pi$
decays, absence in $B^{\pm} \rightarrow K\pi$ decays and predicts unexpected contrast between pure I=1/2
in individual $B^{\pm}$ and $B^o$ final states and
I=1/2 violation in relations between them.

$B(B^o \rightarrow K^+ \pi^-) - 2B(B^o\rightarrow K^o \pi^o) =
 %(19.4 \pm 0.6)- 2\cdot (9.4 \pm 0.6) =
 (0.6 \pm 1.3)\cdot 10^-6   \approx 0$

$2B(B^+ \rightarrow K^+ \pi^o) -
B(B^+ \rightarrow K^o \pi^+ ) =
%(25.8\pm 1.2) - (23.1 \pm 1.0) =
(2.7 \pm 1.6)\cdot 10^-6   \approx 0$

${{\tau^o}\over{\tau^+}}\cdot 2B(B^+ \rightarrow K^+ \pi^o) -
B(B^o \rightarrow K^+ \pi^- ) =
(4.7 \pm 0.82)\cdot 10^-6 \not= 0$

\noindent Analysis of $B\rightarrow K\pi$ data predicts above observed isospin relations and explains dependence on spectator quark flavor.
$B^+ \rightarrow K\pi$ tree diagram $\bar b u\rightarrow \bar s u \bar u u$ has two identical $u$ quarks from weak vertex and spectator. The Pauli principle  requires these quarks at short distances to have wave functions antisymmetric in color or spin. The eigenvalues of conserved symmetries remain entangled in a final state of two separated mesons. This Pauli entanglement suppresses tree-penguin interference and CP violation
in $B^+$ decay but not in $B^o$ decay with spectator $d$ quark.
 The four-body wave function must have two antiquarks with the same symmetry combining with two u-quarks to fragment into a two-pseudoscalar-meson state even under charge conjugation with angular momentum zero. It is classified in the 27-dimensional representation of flavor SU(3) with isospin I=2 for the $\pi \pi$ state and V spin V=2 for the corresponding strange state which is linear combination of $K\pi$ and $K\eta_8$.  These symmetries remain entangled in four-body wave function even after separation into two mesons. Strong Pauli suppression in tree transitions to $K\pi$ which has only a small V=2 component and is mainly V=1. No Pauli suppression in transitions to I=2 $\pi \pi$ state  with also two $u$ quarks but different color-spin couplings.
Standard definition of independent color favored and suppressed tree diagrams
in  $B^\pm\rightarrow K\pi$ decays neglects $uu$ Pauli entanglement.
 %between two diagrams differing by  interchange of two $u$ quarks.

\vspace*{4mm}

\vfill\eject

\section {Introduction - The Three $K-\pi$ Puzzles}

\subsection {Puzzle \# 1. Direct CP Violation observed in
 $B^o \rightarrow K\pi$; not in $B^{\pm} \rightarrow K\pi$ decays}

A general theorem from CPT invariance shows\cite{lipCPT} that direct CP violation can occur only via the interference between two amplitudes which have
different weak phases and different strong phases. This  holds also for all
contributions from new physics beyond the standard model which conserve CPT.

Direct CP violation has been experimentally observed\cite{PDG,HFAG}
in $B_d \rightarrow K^+ \pi^-$ decays.
\beq{acp0}
A_{CP}(B_d \rightarrow K^+ \pi^-)= -0.098 \pm 0.013
\eeq
 The experimental observation (\ref{acp0}) and the knowledge that the penguin
amplitude is dominant for the decay\cite{Ali} require that the decay amplitude must
contain at least one additional amplitude with both weak and strong phases
different from those of the penguin.

The CP violation (\ref{acp0}) has been attributed to the interference between the large contribution from the dominant penguin diagram and smaller contributions from tree diagrams.  The failure to observe CP violation in charged
decays\cite{Ali} has been considered a puzzle\cite{nurosgro,ROSGRO}
because changing the flavor of a spectator quark which does not participate
in the weak decay vertex is not expected to  make a difference.
 \beq{acp+}
\begin{array}{ccl}
\displaystyle
A_{CP}(B^+ \rightarrow K^o_S \pi^+)= 0.009 \pm 0.029
\hfill\\
\\
A_{CP}(B^+ \rightarrow K^+ \pi^o)= 0.051 \pm 0.025
\end{array}
\end{equation}
\subsection {Puzzle \# 2. Experimental relations      not predicted by standard treatments}
New experimental results show that
the individual branching ratios for  $B^o$ and $B^+$ decays agree with
the pure $I=1/2$ amplitude predicted by a penguin diagram.
\beq{acpexp}
\begin{array}{ccl}
\displaystyle
B(B^o \rightarrow K^+ \pi^-) - 2B(B^o\rightarrow K^o \pi^o) =
 (19.4 \pm 0.6)- 2\cdot (9.4 \pm 0.6) = 0.6 \pm 1.3  \approx 0
\hfill\\
2B(B^+ \rightarrow K^+ \pi^o) -
B(B^+ \rightarrow K^o \pi^+ ) = (25.8\pm 1.2) - (23.1 \pm 1.0)
=2.7 \pm 1.6  \approx 0
\end{array}
\eeq
where $B$ denotes the branching ratio in units of $10^-6$.

However, the isospin relation between $B^+ $ and $B^o$ decays predicted by the penguin diagram
is in strong disagreement with experiment.
\beq{newpuz2}
{{\tau^o}\over{\tau^+}}\cdot 2B(B^+ \rightarrow K^+ \pi^o) -
B(B^o \rightarrow K^+ \pi^- ) =
4.7 \pm 0.82 \not= 0
\eeq
where $\tau^o/\tau^+$ denotes the ratio of the $B^o$ and $B^+$ lifetimes and we have used
the experimental values
\beq{newtestex}
\begin{array}{ccl}
\displaystyle
B(B^o \rightarrow K^+ \pi^-)  =
19.4 \pm 0.6
 \hfill\\
\\
\displaystyle
{{\tau^o}\over{\tau^+}}\cdot B(B^+ \rightarrow K^o \pi^+) =
{{(23.1 \pm 1.0)}\over{1.07}}=
21.6\pm 0.93
%{{(23.1 \pm 1.0)}\over{1.07}} - (19.4 \pm 0.6) =
%21.6\pm 0.93 -19.4 \pm 0.6 =
%2.2 \pm 1.1 \propto -\vec P \cdot \vec T
\hfill\\
\\
\displaystyle
{{\tau^o}\over{\tau^+}}\cdot 2B(B^+ \rightarrow K^+ \pi^o)
%-B(B^o \rightarrow K^+ \pi^- )
= {{2\cdot (12.9\pm 0.6)}\over{1.07}}=
%- (19.4 \pm 0.6) =
24.1\pm 0.56
\hfill\\
\\
\displaystyle
%2B(B^o\rightarrow K^o \pi^o) - B(B^o \rightarrow K^+ \pi^-) =
B(B^o\rightarrow K^o \pi^o) =
(9.4 \pm 0.6)
%2\cdot (9.4 \pm 0.6) - (19.4 \pm 0.6) =
%-0.6 \pm 1.3
%\propto - \vec P \cdot (\vec T + \vec S)
\end{array}
\end{equation}

The relation (\ref{newpuz2}) shows  that the $B\rightarrow K\pi$ transition is not a pure penguin.  The relation (\ref{acpexp}) shows  that the $I=1/2$ prediction by a pure  penguin diagram holds for the individual charged and neutral decays and is violated only in the ratio of the branching ratios for charged and neutral  decays.

The significant difference between the experimental values of expressions (\ref{acpexp}) and (\ref{newpuz2})
is not expected in the conventional analyzes. The two relating branching ratios for individual charged and neutral decays still vanish here while one relating charged and neutral case is finite. This indicates a surprising cancelation and motivates a search for a theoretical explanation.

\subsection {Puzzle \# 3.Tree Diagrams for $B^+ \rightarrow K \pi$ and $B^+ \rightarrow \pi \pi$ not related}
Tree diagrams for both $B^+ \rightarrow K \pi$ and $B^+ \rightarrow \pi \pi$ decays have two identical $u$ quarks in final state.
The tree diagram is Pauli blocked in $B^+ \rightarrow K \pi$; experimentally not blocked in $B^+ \rightarrow \pi \pi$

\subsection {The Sum and Difference Rules}
The standard treatment\cite{approxlip,approxgr,Gronau,ketaprimfix} assumes that four $B\rightarrow K\pi$ branching ratios
are determined by three parameters, the dominant penguin diagram $P$  and two interference terms $P\cdot T$ and $P\cdot S$ between the dominant penguin diagram shown in Fig. 1 and the color-favored and color suppressed tree
diagrams shown in Figs. 2 and 3. This treatment assumes the two tree contributions are independent and neglects Pauli blocking. It also assumes that the two tree amplitudes are sufficiently small to be
treated in first order. Second order terms $T\cdot T$, $S\cdot T$
and $S\cdot S$ are assumed to be negligible. These assumptions lead to a sum rule.
\beq{sumruleapp}
R_L \equiv 2{{\Gamma(B^+ \rightarrow K^+ \pi^o) + \Gamma(B^o
\rightarrow K^o \pi^o)} \over {\Gamma(B^+ \rightarrow K^o \pi^+ )
+  \Gamma(B^o
\rightarrow K^+ \pi^-)}} \approx 1
\eeq

The agreement\cite{Ali} with experiment\cite{PDG,HFAG} confirms these assumptions  \cite{approxlip,approxgr,Gronau,ketaprimfix}.

We now investigate what is observable in the experimental data, how to separate
the signal from the noise, how to find the`tree amplitude needed for tree-penguin
interference. We first examine what can be learned from new experimental data.
The sum rule (\ref{sumruleapp}) has been rearranged \cite{ketaprimfix}
to obtain a ``difference rule"
\beq{eqapp}
{{\tau^o}\over{\tau^+}}\cdot \left[ 2B(B^+ \rightarrow K^+ \pi^o)
- B(B^+ \rightarrow K^o \pi^+ )\right] \approx
B(B^o \rightarrow K^+ \pi^-)  - 2B(B^o\rightarrow K^o \pi^o)
\end{equation}
where  the result was expressed in terms of  branching ratios, denote
by B(). This relation (\ref{eqapp}) states that the $I=3/2$ contributions to charged and neutral
decays are equal.

\section {Implications of systematics in the new data}
\subsection {Summary of data requiring explanation}
\begin{enumerate}
\item CP violation in neutral B decays implies tree-penguin interference
\item Absence in charged B decays implies reduced tree-penguin interference
\item Penguin independent of spectator flavor; same in charged and neutral decays
\item Tree contribution depends on spectator flavor
\begin{itemize}
\item In charged B decays tree vertex and spectator both produce $u$ quark.
\item In neutral B decays tree vertex and spectator produce different quark flavors;
\item Possible Pauli suppression in charged B decays;  No possibility In neutral B decays
\end{itemize}
\item Tree
%contributions
in individual charged and nuclear decays both produce I=1/2 final state
\item $B^{\pm}\rightarrow \pi^+\pi^o$ tree
%vertex
and spectator both produce $u$ quark. No observed Pauli suppression
\end{enumerate}

Some of these puzzles can be resolved for $B\rightarrow K\pi$ decays by extreme Pauli suppression; i.e. kill all amplitudes leading to final states
containing two quarks of the same flavor. But  puzzle for $B^{\pm}\rightarrow \pi^+\pi^o$ decays is sharpened.
\begin{enumerate}
\item No tree contribution to charged B decays; no CP violation
\item Diquark produced in neutral decays must be $ud$; no $uu$ diquark allowed
\item $ud$ diquark produced in neutral decays has I=0; final state has I=1/2
\item Creates puzzle In $B^{\pm}\rightarrow \pi^+\pi^o$ where $uu$ diquark needed for observed tree contribution
\end{enumerate}
In the remainder of this paper we first show how all puzzles are resolved by a proper treatment of the Pauli principle, neglected in previous treatments\cite{nurosgro,ROSGRO} including the spin and color degrees of freedom.
We then show how the Pauli antisymmetry relates the two tree amplitudes called color favored and color suppressed, which have been considered independent\cite{nurosgro,ROSGRO}. Finally we show how a flavor-topology analysis generalizes this approach to include final state interactions.
\subsection {SU(3) breaking prevents using parameters from $B\rightarrow \pi\pi$
decays for $B\rightarrow K\pi$}
Standard treatments \cite{nurosgro,ROSGRO} of charmless B decays have used data from
$B\rightarrow \pi\pi$ decays together with SU(3) flavor symmetry to obtain parameters
for analysis of $B\rightarrow K\pi$. At that time precise $B\rightarrow K\pi$ data were not yet
available. New more precise data revealed contradictions with this approach\cite{nuhuor1}.
The source of these contradictions can be seen as due to SU(3) breaking.

 A final $\pi^o\pi^+$ state $\ket{f;\pi^o\pi^+}$ is a pure $I=2$ state in a pure SU(3)
 27-dimensional representation of flavor SU(3).

In the symmetry limit the strange analog of the $\pi^o\pi^+$ state in a pure SU(3)
27 is $K^+ V_{10}$ state where $V_{10}$ denotes the $V$ spin analog of the $\pi^o$
with $V=1,V_z=0 $. This state is badly broken by SU(3) symmetry breaking into
$K^+ \pi^o$, $K^+ \eta$ and $K^+ \eta'$. The $K^+ \pi^o$ state has only (1/4) of the
SU(3) 27 related to the $B\rightarrow \pi\pi$ decay. The remaining (3/4) is classified in
other representations of SU(3) which are not related to the $B\rightarrow \pi\pi$ decay.
Thus there is no possibility for using SU(3) with $B\rightarrow \pi\pi$ decay to
obtain parameters for analysis of $B\rightarrow K\pi$.

\subsection{An approximate quantitative treatment of Pauli effects in $B \rightarrow K \pi$ decays.}

The tree diagram for the transition from an initial $B$ meson state  consisting of a $\bar b$ antiquark and a nonstrange spectator quark
to a strange charmless two-meson final state is written
\beq{secquanx}
\begin{array}{ccl}
\displaystyle
\ket{B_d}=\bar b d \rightarrow
\bar s  \cdot \left[u\bar u \right] d
%= \bar s  \cdot \left[\kappa d \bar d+ (1+\xi)\cdot  ud \bar u \right] d\approx
%\bar s  \cdot (1+\xi)\cdot  ud \bar u
\hfill\\
%u\\
\ket{B_u}=\bar b u \rightarrow
%\bar s  \cdot \left[ \cdot u\bar u \right] u=
\bar s  \cdot \left[ \kappa\cdot  u\bar u \right] u
%\approx \bar s  \cdot du \bar d
\end{array}
\end{equation}
where the parameter $\kappa$ is a Pauli factor expressing the probability that the two u quarks are not Pauli blocked because they are not  in the same color-spin state.

We first consider the approximation $\kappa \approx 0$ where a $u$ quark produced by a weak interaction cannot enter the same state as a
$u$ spectator quark.
The states with $\kappa=0$ have no quark pairs of the same flavor. We call these Pauli-favored states.

When $\kappa=0$ the tree diagram is finite for neutral decays but vanishes in charged decays. This solves Puzzle \#1 by suppressing the tree contribution and CP
violation in  charged B decay while allowing the tree contribution in neutral decays.
This suppression is lost in conventional treatments which consider color-favored and color-suppressed
tree amplitudes as independent without considering Pauli suppression.

The tree amplitude produces a $u \bar u$ pair in the $b$ decay vertex. This amplitude is Pauli suppressed in charged $B$ decays where the spectator quark is also a $u$ quark. Thus in the $\kappa = 0$ approximation tree-penguin interference which might produce CP violation is present in neutral decays and absent in charged decays. This can explain how CP violation can be drastically changed by changing the spectator quark
and the otherwise mysterious result  (\ref{acp+}). We now go beyond the $\kappa =0$ approximation and consider a full color-spin analysis.

\section {Analysis with Pauli entanglement and color-spin}
\subsection {Detailed symmetry and Pauli analysis}
The puzzles can  hopefully be resolved by a complete QCD calculation which is not yet feasible. We look for symmetry methods which can give results without such QCD calculations. We first consider the symmetries of the four-body $ qq\bar q\bar q$ state a very short time after the decay when they are still in a very small region of configuration space. We assume that for all subsequent times angular momentum, isospin, flavor SU(3), color SU(3) gauge theory  and charge conjugation invariance are preserved. The symmetry quantum numbers of the state at short times are assumed to be  preserved at longer times with entanglement if necessary. There is no need to consider the  difference between the spatial wave functions of the spectator b quark and the recoiling light meson. Their symmetries are entangled
even at large distances.

The dependence on spectator flavor arises from the Pauli blocking
by the spectator quark of a quark of the same flavor participating in the weak vertex. The u-quark produced by a
tree diagram is Pauli blocked by the spectator $u$ quark in $B^+$ decay but
is not affected by the spectator $d$ quark in neutral decays. This difference in Pauli
blocking suppresses the tree contribution and CP violation in charged $B$ decays but allows
tree-penguin interference and enables CP violation to be observed in neutral decays.

For a quantitative treatment of Pauli blocking we first note that the tree diagram for the decay of a $\bar b$ antiquark to a charmless final state is described by the
vertex
\beq{Bvert}
\bar b \rightarrow
\bar q u \bar u
\end{equation}
where $\bar q$ denotes a  $\bar d$ antiquark for $\pi\pi$ decays or a
$\bar  s$ antiquark for $K\pi$ decays.

Symmetry restrictions from the Pauli principle arise when a nonstrange spectator
quark has the same flavor as the $u$ quark emitted from the weak vertex. This occurs
in the tree diagram for $B^+$ where the final state contains two $u$ quarks.
\beq{tree}
B^+ =\bar b u \rightarrow
\bar q u \bar u u
\end{equation}

In B decays to two pseudoscalar mesons a spin-zero state decays into two spin-zero particles with zero
internal orbital angular momentum. To conserve angular momentum the final state must
have no orbital angular momentum.
A flavor-symmetric $uu$ state in a spatially symmetric S-wave is required by the Pauli
principle to be
antisymmetric in color or spin.
The  antiquark  pair in (\ref{Bvert}) must also be antisymmetric in either color or spin.
Although no Pauli principle forbids a symmetric color - spin state
such states cannot combine with the $uu$ pair to make the spin-zero color singlet
final state $\pi\pi$ or $K\pi$.
The
fragmentation of a $uu\bar u \bar q$ state into a
$\pi^+\pi^o$ or $K^+\pi^o$ is a strong interaction which conserves flavor SU(3) and
charge conjugation.

Both the $uu$ diquark and the $\bar u \bar q$ antidiquark are thus
antisymmetric in color or  spin. The generalized Pauli principle requires each
to be symmetric in flavor SU(3) and its  SU(2) subgroup isospin for $\pi\pi$ decays or
V-spin for $K\pi$ decays. Each is therefore respectively in
the symmetric isospin state with $I=1$ or in the symmetric V-spin state with $V=1$

The pion isotriplet has isospin one and odd G parity.
The isoscalar pseudoscalar mesons $\eta$ and $\eta'$ have isospin zero and even G parity.
A nonstrange final state must be even under  charge conjugation  and have even
G parity to decay into two pions % pseudoscalar mesons
in an orbital S wave.
Dothan\cite{dothan}  generalized the idea of G parity from SU(2) to SU(3).
We call the generalization of G parity to SU(3) and SU(n)``Dothan parity".
For SU(3) Dothan parity
defines the relative phases of the charge conjugate states in the same SU(3) octet
and defines the eigenvalue under charge conjugation of its C-eigenstate members.
We denote the the V- spin (us) analog of G parity by $G_V$.
The $K^+$ and the three members of the V spin triplet have V spin 1 and odd $G_V$ parity.
The V-spin scalar and  vector pseudoscalar mesons are linear combinations of $\pi^o$, $\eta$ and $\eta'$
with V-spin zero, even $G_V$ parity and with V-spin one, odd $G_V$ parity. The $\pi^o$ is (3/4)
V-spin zero, even $G_V$ parity and (1/4) with V-spin one, odd $G_V$ parity.

To produce a final state with even $G$ parity
the $(I=1,I_z=+1)$ diquark and the $(I=1,I_z=0)$ antidiquark must be coupled
symmetrically to  $(I=2,I_z=+1)$.
Similarly the $(V=1,V_z=+1)$ diquark and the $(V=1,V_z=0)$ antidiquark must be coupled
symmetrically to  $(V=2,V_z=+1)$ to produce a final state with even $G_V$ parity
.
These states are in the 27-dimensional representation of flavor SU(3).

The final states in the 27 are produced from a $u$ quark pair in a color-spin state which
satisfies the Pauli Principle. Final states of two pseudoscalar mesons in other representations
of $SU(3)_{flavor}$ are Pauli suppressed.

A final $\pi^o\pi^+$ state $\ket{f;\pi^o\pi^+}$ is a pure $I=2$ state in a pure SU(3)
27.
Thus the tree diagram for the nonstrange transition
$(B^+ \rightarrow \pi+ \pi^o)$ is not Pauli suppressed.

A final $K^o\pi^+$ state $\ket{f;\pi^oK^+}$
has no $V=2$ component, since both the $K^o$ and $\pi^+$
have V=1/2. Thus the tree diagram for the $K^o\pi^+$ decay must vanish and this
decay is pure penguin.

The final $K^+\pi^o$ state contains a $\pi^o$ which is a linear combination of
$V=0$ and $V=1$ states with probability of 1/4 for $V=1$.
The component with $V=0$ cannot combine with a $V=1$ $K^+$ to make $V=2$.
The $V=1$ component can   combine with a $V=1$ to make $V=2$  with a
probability of 1/2.
Thus the probability that the final
$K^+\pi^o$ state has a $V=2$ component is 1/8.
Thus we see that Pauli blocking suppresses the tree diagram for the
$(B^+ \rightarrow K^+ \pi^o)$
transition by a factor 8.

Present data are consistent with complete suppression but evidence for a
partial suppression is still down in the noise.

The $ud\bar u \bar s$ state created in the tree diagram for
$B_d$ decay has no such restrictions. It can be in a flavor SU(3)
octet as well as a 27. Its ``diquark-antidiquark" configuration includes the
flavor-SU(3) octet constructed from the spin-zero color-antitriplet
flavor-antitriplet ``good" diquark found in the  $\Lambda$ baryon and its
conjugate ``good" antidiquark. These ``good diquarks" do not exist in the
corresponding $uu\bar u \bar s$ configuration.

We again see that the Pauli effects produce a drastic dependence on spectator
quark flavor in the tree diagrams for $B \rightarrow K \pi$ decays.
Tree-penguin interference can explain both the presence
of CP violation in neutral decays and its absence charged decays.

We now note that the $ud$ pair in the final states must be isoscalar by the generalized Pauli principle. The final states must then be pure isospin eigenstates with $I=1/2$ and confirm the experimental result (\ref{acpexp}). In the standard treatments\cite{nurosgro,ROSGRO} the $I=3/2$ component is not suppressed in pure tree transitions

\section {Conventional Analysis using new data without Pauli suppression}

We now investigate a detailed conventional analysis of new experimental data with no new theory.
We later show how the experimentally observed cancelation between the results (\ref{acpexp} and (\ref{newpuz2}) can arise from the Pauli principle.

\subsection{Four independent measurements determined by three parameters}

Four experimental branching ratios
for  $B \rightarrow K\pi$ are available\cite{PDG,HFAG}.
The conventional analysis expresses their four
amplitudes in terms of three amplitudes\cite{approxlip,approxgr,Gronau,ketaprimfix}.
\begin {enumerate}
\item The gluonic penguin diagram, denoted by $P$ and shown in Fig. 1
\item The color-favored tree diagram, denoted by $T$ and shown in Fig. 2
\item The color-suppressed tree diagram, denoted by $S$ and shown in Fig. 3
\end {enumerate}
\beq{pst}
\begin{array}{ccl}
\displaystyle
A[K^o\pi^+]=P; ~ ~ ~ A[K^+\pi^-]= T + P
\hfill\\
\\
\displaystyle
A[K^o\pi^o]={{1}\over{\sqrt{2}}} [S -  P]; ~ ~ ~
A[K^+\pi^o]={{1}\over{\sqrt{2}}} [T + S + P]
\end{array}
\end{equation}
The standard treatment\cite{approxlip,approxgr,Gronau,ketaprimfix} neglects Pauli blocking and assumes that  the two tree contributions are independent and are sufficiently small to enable
the interference terms to taken only to first order,

\beq{pstsq}
\begin{array}{ccl}
\displaystyle
|A[K^o\pi^+]|^2
=|\vec {{ P}}|^2; ~ ~ ~ |A[K^+\pi^-]|^2 \approx |\vec {{ P}}|^2+2\vec { P} \cdot \vec T
\hfill\\
\\
\displaystyle
2\cdot |A[K^o\pi^o]|^2
\approx |\vec {{ P}}|^2  -\vec  P \cdot \vec S; ~ ~ ~
2\cdot |A[K^+\pi^o]|^2
\approx |\vec {{ P}}|^2+2\vec { P} \cdot (\vec T+\vec S)
\end{array}
\end{equation}
where the approximate equalities hold to first order in the $T$ and $S$
amplitudes.

\subsection {Conventional analysis of the difference rule}

Three different independent
differences between these branching ratios can be defined which eliminate the
penguin contribution. We choose the expressions
(\ref{acpexp}) and (\ref{newpuz2}) and express them in terms of the three parameters.
 \beq{acpexpc}
 \begin{array}{ccl}
\displaystyle
B(B^o \rightarrow K^+ \pi^-) - 2B(B^o\rightarrow K^o \pi^o) =
\vec { P} \cdot \vec T -\vec  P \cdot \vec S
 %(19.4 \pm 0.6)- 2\cdot (9.4 \pm 0.6)
 = 0.6 \pm 1.3  \approx 0
\hfill\\
2B(B^+ \rightarrow K^+ \pi^o) - B(B^+ \rightarrow K^o \pi^+ ) =
2\vec { P} \cdot (\vec T+\vec S)
%(25.8\pm 1.2) - (23.1 \pm 1.0)
=2.7 \pm 1.6  \approx 0
\hfill\\
{{\tau^o}\over{\tau^+}}\cdot 2B(B^+ \rightarrow K^+ \pi^o) -
B(B^o \rightarrow K^+ \pi^- ) =
{{\tau^o}\over{\tau^+}}\cdot 2\vec { P} \cdot (\vec T+\vec S)
2\vec { P} \cdot \vec T
=4.7 \pm 0.82 \not= 0
 \end{array}
\eeq

The relation (\ref{eqapp}) states that the $I=3/2$ contributions to charged and neutral
decays are equal.
Combining this result (\ref{eqapp}) with the approximate experimental result (\ref{newpuz})
shows that in this approximation the $I=3/2$ contributions to
both charged and neutral decays vanish.

Combining the equations (\ref{acpexpc})gives the relation:
\beq{newtest0}
\frac{\vec P\cdot (\vec T + \vec S)}{\vec P\cdot (\vec T - \vec S)}
=\frac{2B(B^o\rightarrow K^o \pi^o) - B(B^o \rightarrow K^+ \pi^-)} {{{[\tau^o}/{\tau^+]}}\cdot[ B(B^+ \rightarrow K^o \pi^+) + 2B(B^+ \rightarrow K^+ \pi^o)] -
2B(B^o \rightarrow K^+ \pi^- )}
=0.09\pm 0.1
\end{equation}

The new precise data show  that
the interference term between the dominant penguin amplitude
and the color-suppressed tree amplitude $\vec P \cdot \vec S $is definitely
finite and well above the experimental errors. The sum
rule is still satisfied within two standard deviations and is now nontrivial.
But the interference term
$\vec P \cdot (\vec T+\vec S)$ is now equal to zero
well within the experimental errors (\ref{newtest0}). This confirms
the Pauli symmetry prediction.

There is no new theory here.
Choosing three independent differences in a way to minimize experimental errors
shows significant signals  well
above the noise of experimental errors that still fit an
overdetermination of the two parameters and lead to the result (\ref{newtest0}). The relations
between charged and neutral decays show two finite
tree-penguin interference contributions that can produce the observed
direct CP violation in neutral B-decays.  However the third difference between two
neutral decays  is
consistent with the pure penguin prediction of zero well below the noise and below the other two
contributions. The absence of tree-penguin contributions in this difference is
completely unpredicted in the standard treatments.

The two transitions (\ref{secquanx})  which have a d-quark spectator are described
respectively by the color-favored and color-suppressed tree diagrams shown respectively in
figs. 2 and 3 in addition to the dominant common penguin diagram  shown in
fig. 1 . This
cancelation between the contributions of the two tree diagrams is surprising because
the standard treatments assume that the these two tree contributions
are completely independent and are not expected to cancel.

\subsection{The surprising cancelation suggests Pauli effects $P\cdot (T + S) \approx 0 $}

The Pauli principle  neglected  in conventional treatments
can produce the cancelation (\ref{newtest0}). The amplitudes $T$ and $S$  go into one another under
the interchange of the two identical $u$ quarks in $A[K^+\pi^o]$.
A full examination
of Pauli effects requires antisymmetrization of the $uu$ wave function including the
color and spin correlations. As a  first approximation we neglect color and spin.
Then Pauli antisymmetry
requires $T$ and $S$ amplitudes to be equal and opposite and explains the cancelation
(\ref{newtest0}).
Both $B^+$ decays are then pure penguin decays to the $I=1/2$ $K\pi$ state.
Experiment\cite{HFAG} shows  agreement with this prediction to  between one
and two standard deviations.

Thus tree-penguin interference with normally ignored Pauli effects
can explain the observed CP violation in
charged B-decays and its absence in neutral decays.

This shows how a nontrivial change in the weak
decay amplitude can arise from a change of the flavor of the spectator quark.

\subsection{Symmetry arguments supporting the vanishing of $\vec  P \cdot (\vec T + \vec S)$}

In the $u n \bar u \bar s$ tree diagrams for charmless strange $B$ decays, the $\bar b \rightarrow u \bar u \bar s$ transition produces a four-quark state $u n \bar u \bar s$ where $n$ denotes the nonstrange spectator quark.

The $K^+\pi^-$ final state $(u  \bar s)(n \bar u)$ can be produced by a $B^o$ tree diagram in which the spectator quark $n$ is a $d$ quark and combines with the $\bar u$ in a color-favored transition shown in Fig. 2.
The CP violation observed in this state indicates that it is produced by appreciable $P\cdot T$ interference.

The $K^+\pi^o$ final state $(u  \bar s)(n \bar u)$ or $(n  \bar s)(u  \bar u)$ can be produced by a $B^+$ tree diagram in which the spectator quark $n$ is a $u$ quark  and combines with either the $\bar u$ in a color-favored transition
shown in Fig. 2
or the $\bar s$ in a color-suppressed transition shown in Fig.3. The failure to observe CP violation in this state while CP violation is observed in the $K^+\pi^-$ final state indicates that both $P\cdot T$
and $P\cdot S$ interference contributions are appreciable and their interference contributions have opposite phase and tend to cancel any CP violation.

The experimental data for these two transitions thus present predictions for the following two final states.

The $K^o\pi^o$ final state $(n  \bar s)(n \bar u)$  can be produced by a $B^o$ tree diagram  in which the spectator quark $n$ is a $d$ quark and combines with the $\bar s$ in a color-suppressed transition shown in Fig. 3. This transition is produced by $P\cdot S$ interference  which is expected to be similar to the  $P\cdot T$
interference contribution and produce a similar CP violation to that  observed in
the $K^+\pi^-$ final state

The $K^o\pi^+$ final state $(d\bar s)(u\bar d)$ contains a $\bar d$ antiquark and cannot be produced by a tree diagram leading to a $u n \bar u \bar s$ state.
The prediction for the transition to
this final state is that it has no tree contribution, no penguin-tree interference and no CP violation

%\section {A detailed analysis of the new experimental $B \rightarrow K\pi$ data}

\section{A flavor topology analysis which includes final state interactions}
The unique flavor topology of the charmless strange quasi-two-body weak B decays
enables the results (\ref{newtest0}) to be obtained in a more general analysis
of these decays
including almost all possible diagrams including final state interactions and complicated
multiparticle intermediate states.

Consider diagrams for a charmless $B(\bar b q_s)$ decay into one strange and
one nonstrange meson, where $q_s$ denotes either a $u$ or $d$. The allowed
final  states must have the quark constituents  $\bar s n \bar n q_s$ where $n$
denotes a $u$ or $d$ nonstrange quark. We consider the topologies of all
possible diagrams in which a $\bar b$ antiquark and a nonstrange quark enter a
black box from which two final $q \bar q$ pairs emerge. We follow the quark
lines of the four  final state particles through the diagram going backward and
forward in time until they reach either the initial state or a vertex where
they are created. There are only two possible quark-line topologies for these
diagrams:

\begin{enumerate}

\item We call a generalized penguin diagram, shown in Fig. 1 , the sum of all possible diagrams in which a $\bar q q$ pair appearing in
the final state  is created by a gluon somewhere in the diagram. The quark
lines for the  remaining pair must go back to the weak vertex or the initial
state. This diagram includes not only the normally called penguin diagram but
all other diagrams in which the final  pair is created by gluons somewhere in
the diagram. This includes for example all diagrams normally called ``tree
diagrams" in which an outgoing $u \bar u$ or $c \bar c$ pair  is annihilated into gluons in a
final state interaction and a new isoscalar  $\bar q q$ pair is created by the
gluons.
There are two topologies for penguin diagrams.
\begin{itemize}
\item. A normal penguin diagram has a the spectator quark line continuing unbroken
 from the initial state to the final state.
 This  penguin contribution is described by a single parameter,
denoted by $P$ which is independent of the spectator quark flavor and contributes
equally to the $\bar s u \bar u q_s$ and $\bar s d \bar d q_s$ states.
\item A diagram which we call here an ``anomalous penguin" has the spectator
``u" quark in a $B^+$ decay annihilated in a final state interaction
against the $\bar u$ antiquark produced in a tree diagram. This diagram also   contributes
equally to the $\bar s u \bar u q_s$ and $\bar s d \bar d q_s$ states. But this
diagram denoted by $P_u$ is present only in
charged decays.
\end{itemize}
\item We call the ``tree diagram" the sum of all possible diagrams in which all
of the four quark lines leading to the final state go back to a initial $\bar s
u \bar u$ state created by the weak decay of the $b$ quark and the $ q_s$
spectator whose line goes back to the initial state.
     There are two possible couplings of the pairs to create final two meson
states from this diagram
 \begin{itemize}

\item The $\bar s u$ pair is coupled to make a strange meson; the  $\bar u q_s$
pair is coupled to make a nonstrange meson as shown in Fig.2. This is
conventionally called the  color-favored coupling. The contribution
of this coupling is described by a  single parameter, denoted by $T$.

\item The $\bar s q_s$ pair is coupled to make a strange meson; the  $u \bar u$
is coupled to make a nonstrange meson as shown in Fig. 3.
This is conventionally called the  color-suppressed coupling. The contribution
of this coupling is described by a  single parameter, denoted by $S$.

 \end{itemize}
 \end{enumerate}

All the results (\ref{newtest0}) obtained with the conventional definitions of $P$, $T$ and $S$ are seen to hold here
with the new definitions of $P$, $T$ and $S$. They now include contributions from
all final state interactions which conserve isospin and do not change
quark flavor.
The one final
state interaction not included is the $P_u$ diagram occurring in $B^+$ decays.
% where the spectator $u$ quark
%annihilates a $\bar u$ antiquark into color singlet gluons and creates an isospin
%singlet $n \bar n$.
The flavor topology of this diagram creates an additional $I=1/2$ state
which is neglected in the derivation of the results (\ref{newtest0}).
These results hold as long  as the contribution of this $P_u$ diagram by final state interactions to the observed final states is negligible.

The additional $I=1/2$ contribution does not affect the ``difference rule" (\ref{eqapp}) which considers only the $I=3/2$ contributions.

In neutral $B_d$ decays there is no $P_u$ diagram.
Thus the simple relations (\ref{newtest0}) between the
$P$, $T$ and $S$ amplitudes hold for neutral decays are
%the same as in the simple model and
not changed by isospin conserving final state interactions.

Further analysis of the contribution of this additional $I=1/2$ contribution is needed to include its modification of tree-penguin interference in obtaining definite values for CP violation in charged $B$ decay.

The electromagnetic penguin diagram is also  included in this flavor-topology formulation. The photon coupling to a $q \bar q$ pair can can be written
\beq{photon}
\gamma \rightarrow 2 u \bar u - d \bar d - s\bar s = 3u \bar u - [u \bar u + d \bar d + s\bar s]
\eeq
This coupling is included in the flavor-topology formulation as a linear combination of a tree coupling and a penguin coupling and contributes to the results (\ref{newtest0}). However there is now no simple relation between the $P$, $T$ and $S$ amplitudes and CKM matrices.

\section {Comparison with other approaches}

Previous analyses \cite{nurosgro,ROSGRO} were performed at a time when experimental
values for $B\rightarrow K\pi$ branching ratios were not
sufficiently precise to enable a significant test of the sum rule
(\ref{eqapprev}). Values of each of the three interference terms in (\ref{newtest0})
were statistically consistent with zero.
The full analysis required the use of data from $B\rightarrow \pi\pi$ decays and
the assumption of $SU(3)_{flavor}$ symmetry. Contributions of the electromagnetic
penguin diagram were included and the relevant CKM matrix elements were included. But
there was no inclusion of constraints from the Pauli principle nor contributions from
final state interactions.

The present analysis uses new experimental data which enable a statistically significant
evaluation of the interference terms (\ref{newtest0}) without additional information
from $B\rightarrow \pi\pi$ decays or the assumption of $SU(3)_{flavor}$ symmetry.
Constraints from the Pauli principle are included in a general calculation including
color and spin and entanglement. Symmetries of the original weak amplitude are preserved  with
entaglement in the final state of two separated  mesons. Contributions from all isospin invariant finite state
interactions are included as well as constraints from the Pauli principle. The
flavor topology definition of the interference parameters includes contributions from the electromagnetic penguin diagram since the  quark states in final state of a photon can be rewritten as the sum of an isoscalar and
a $u\bar u$ state. However the flavor topology parameters are no longer simply related to the
CKM matrix elements. Additional assumptions and information are necessary to determine the CKM matrix elements
and explain CP violation.

The main advantage of this approach is that it gives simple explanations for the  absence of CP violation (\ref{acp+}) in charged B decays, the observed absence of an $I=3/2$ component in the final state,
and the vanishing of the experimental value
 (\ref{newtest0})

This vanishing of tree-penguin interference $B^+$ decays is explained by a symmetry analysis including the constraints of the Pauli principle and entanglement on states containing a pair of identical $u$ quarks.
\section {Conclusion}

Experiment has shown that the penguin-tree interference contribution in $B^+
\rightarrow K^+\pi^o$ decay is very small and may even vanish. The
corresponding interference contributions to neutral $B\rightarrow K\pi$ decays
have been shown experimentally to be finite. In charged decays the
previously neglected Pauli antisymmetrization  produces a cancelation
between color-favored and color-suppressed tree diagrams which differ by the
exchange of identical $u$ quarks. This explains the smallness of penguin-tree interference and
small CP violation in charged $B$ decays. Pauli cancelation does not occur in neutral decay diagrams
which have no pair of identical quarks.
 This can explain why CP violation
has been observed in neutral $B \rightarrow K\pi$ decays and not in charged
decays
\section*{Acknowledgements}

This research was supported in part by the U.S. Department of Energy, Division
of High Energy Physics, Contract DE-AC02-06CH11357. It is a pleasure to thank
Michael Gronau, Yuval Grossman, Marek Karliner, Zoltan Ligeti, Yosef Nir,
Jonathan Rosner, J.G. Smith, and  Frank Wuerthwein for discussions and
comments.

%----------------------------------------------------------------------
% This prevents REFERENCES from forcing a page break
%\def\newpage{\vskip10ex}
%
\catcode`\@=11 % This allows us to modify PLAIN macros
\def\references{
\ifpreprintsty \vskip 10ex
%\ifpreprintsty \newpage
%
\hbox to\hsize{\hss \large \refname \hss }\else
\vskip 24pt \hrule width\hsize \relax \vskip 1.6cm \fi \list
{\@biblabel {\arabic {enumiv}}}
{\labelwidth \WidestRefLabelThusFar \labelsep 4pt \leftmargin \labelwidth
\advance \leftmargin \labelsep \ifdim \baselinestretch pt>1 pt
\parsep 4pt\relax \else \parsep 0pt\relax \fi \itemsep \parsep \usecounter
{enumiv}\let \p@enumiv \@empty \def \theenumiv {\arabic {enumiv}}}
\let \newblock \relax \sloppy
 \clubpenalty 4000\widowpenalty 4000 \sfcode `\.=1000\relax \ifpreprintsty
\else \small \fi}
\catcode`\@=12 % at signs are no longer letters
%-----------------------------------------------------------------

%%%%%%%%%%%%%%%%%%%%%%%%%%%%%%%%%%%%%%%%%%%%%%%%%%

%\tighten
%\preprint {\vbox{
% \hbox{WIS/18 /00-Sept.-DPP}
% \hbox{TAUP 2645-2000}
% \hbox{ANL-HEP-PR-00-100}
% \hbox{hep-ph/0009241}
% \hbox{}
%}}

{\begin{figure}[htb]
$$\beginpicture
\setcoordinatesystem units <\tdim,\tdim>
\stpltsmbl
\putrule from -25 -30 to 50 -30
\putrule from -25 -30 to -25 30
\putrule from -25 30 to 50 30
\putrule from 50 -30 to 50 30
\plot -25 -20 -50 -20 /
\plot -25 20 -50 20 /
\plot 50 20 120 40 /
\plot 50 -20 120 -40 /
\springru 50 0 *3 /
\plot 120 20 90 0 120 -20 /
\put {$\overline{b}$} [b] at -50 25
\put {${q_s}$} [t] at -50 -25
\put {$\overline{s}$} [l] at 125 40
\put {$\overline{n}$} [l] at 125 -20
\put {$n$} [l] at 125 20
\put {${q_s}$} [l] at 125 -40
\put {$\Biggr\}$ $K(\vec k)$} [l] at 135 30
\put {$\Biggr\}$  $\pi(-\vec k)$} [l] at 135 -30
\put {$G$} [t] at 70 -5
\setshadegrid span <1.5\unitlength>
\hshade -30 -25 50 30 -25 50 /
\linethickness=0pt
\putrule from 0 0 to 0 60
\endpicture$$
\caption{\label{fig-2}} \hfill ``Gluonic penguin'' ($P$) diagram.
$G$ denotes any number of
gluons. $n$ denotes $u$ or $d$ quark.
\hfill~ \end{figure}}

{\begin{figure}[htb]
$$\beginpicture
\setcoordinatesystem units <\tdim,\tdim>
\stpltsmbl
\putrule from -25 -30 to 50 -30
\putrule from -25 -30 to -25 30
\putrule from -25 30 to 50 30
\putrule from 50 -30 to 50 30
\plot -25 -20 -50 -20 /
\plot -25 20 -50 20 /
\plot 50 0 120 -20 /
\plot 50 -20 120 -40 /
\photonru 50 20 *3 /
\plot 120 40 90 20 120 20 /
\put {$\overline{b}$} [b] at -50 25
\put {$q_s$} [t] at -50 -25
\put {$\overline{s}$} [l] at 125 40
\put {$u$} [l] at 125 20
\put {$\overline{u}$} [l] at 125 -20
\put {$q_s$} [l] at 125 -40
\put {$\Biggr\}$  $K(\vec k)$} [l] at 135 30
\put {$\Biggr\}$  $\pi(-\vec k)$}  [l] at 135 -30
\put {$W$} [t] at 70 15
\setshadegrid span <1.5\unitlength>
\hshade -30 -25 50 30 -25 50 /
\linethickness=0pt
\putrule from 0 0 to 0 60
\endpicture$$
\caption{\label{fig-4}} \hfill Color favored tree ($T$) diagram.
 \hfill~ \end{figure}}

{\begin{figure}[htb]
$$\beginpicture
\setcoordinatesystem units <\tdim,\tdim>
\stpltsmbl
\putrule from -25 -30 to 50 -30
\putrule from -25 -30 to -25 30
\putrule from -25 30 to 50 30
\putrule from 50 -30 to 50 30
\plot -25 -20 -50 -20 /
\plot -25 20 -50 20 /
\plot 50 20 120 40 /
\plot 50 -20 120 -40 /
\photonru 50 0 *3 /
\plot 120 20 90 0 120 -20 /
\put {$\overline{b}$} [b] at -50 25
\put {${q_s}$} [t] at -50 -25
\put {$\overline{u}$} [l] at 125 40
\put {$u$} [l] at 125 20
\put {$\overline{s}$} [l] at 125 -20
\put {${q_s}$} [l] at 125 -40
\put {$\Biggr\}$ $\pi(\vec k)$}  [l] at 135 30
\put {$\Biggr\}$ $K(-\vec k)$} [l] at 135 -30
\put {$W$} [t] at 70 -5
\setshadegrid span <1.5\unitlength>
\hshade -30 -25 50 30 -25 50 /
\linethickness=0pt
\putrule from 0 0 to 0 60
\endpicture$$
\caption{\label{fig-5}} \hfill Color suppressed tree ($S$) diagram.
\hfill~ \end{figure}}

\end{document}